\documentclass[aps,pre,preprint,superscriptaddress,showpacs,preprintnumbers,amsmath,amssymb]{revtex4}

\usepackage{graphicx}
\usepackage{bm}% bold math
\usepackage{color}
\usepackage{amsmath}

\def\Vect#1{\mbox{\boldmath $#1$}}
\def\cl{{\cal L}}

\def\bs#1{\boldsymbol{#1}}
\begin{document}

\title{Nonequilibrium identities of granular vibrating beds}

\author{Hisao Hayakawa}
\email[]{hisao@yukawa.kyoto-u.ac.jp}
\affiliation{Yukawa Institute for Theoretical Physics, Kyoto University, Kyoto 606-8502, Japan}

%\publishedin{%         %Write this ONLY in cases of addenda and errata
%Prog.~Theor.~Phys.\ \textbf{XX} (19YY), page.}

%\recdate{Mmmmm DD, YYYY}%            %Editorial Office will fill in this.

\begin{abstract}
%{%         %this abstract is neglected when [addenda] or [errata]
We derive the integral fluctuation theorem around a nonequilibrium stationary state for frictionless and soft core granular particles under an external vibration achieved by a balance between 
an external vibration and inelastic collisions.
We also derive the standard fluctuation theorem and the generalized Green-Kubo formula for this system.
%We verify the validity of the derived formulae in terms of the molecular dynamics simulation.

\end{abstract}
%\date{\today}

%\pacs{}

\maketitle

\section{Introduction}

%Construction of a nonlinear response theory around a nonequilibrium steady states is one of the
%most challenging problems in theoretical physics~\cite{Zubarev74,McLennan88,Evans08}.
One of the most remarkable achievements in recent nonequilibrium statitistical mechanics is to demonstrate the existence of some nonequilibrium identities such as the generalized Green-Kubo relation~\cite{Evans08,Morriss87},
various fluctuation theorems~\cite{FT93,GC95,Kurchan,Evans02,Seifert12} and the Jarzynski equality\cite{Jarzynski} as well as the mutual relationship\cite{Crooks}.
These identities are exact and reproduce the conventional Green-Kubo formula, the second law of thermodynamics and Onsager's reciprocal relation in  specific limits.
Therefore, these identities are regarded as fundamental relations in nonequilibrium statistical mechanics.

Although it has been believed that these identities are supported by the local time-reversal symmetry or the detailed balance condition,
some experiments suggest the existence of fluctuation theorem or related equations even in granular systems which do not have any time reversal symmetry\cite{mennon04,Chong09b,kumar11,joubaud12,naert12,mounier12}, though there exists a counter argument \cite{Puglisi05}.
It is remarkable that Puglisi and his coworkers\cite{Puglisi05EPL,Puglisi06PRE,Puglisi06JSM,Sarracino10} clarified that granular fluids do not hold the conventional fluctuation theorem  but have only the second type fluctuation theorem by Evans and Searles~\cite{Evans94}.
As long as the author's knowledge, however, there is only a paper by Chong et al. which has proven the existence of 
both the generalized Green-Kubo relation and the integral fluctuation theorem\cite{Seifert12} for a granular system under a steady plane shear~\cite{Chong09b}.
They also developed the representation of a nonequilbrium steady-state distribution function~\cite{Chong10} and the liquid theory for sheared dense granular systems.\cite{Hayakawa10a}.
%Their paper is a natural extension of a representation of the nonequilibrium steady-state distribution function in terms of the excess entropyproduction developed by~\cite{Komatsu08,Komatsu09}.
Recently, Hayakawa and Otsuki~\cite{HO12} extended their previous formulation to discuss nonequilibrium identities around a nonequilibrium steady state, and also demonstrate their  validities from the direct comparison between the obtained identities and the numerical simulations. 
%They also discussed the effect of measurement. 

Unfortunately, some parts of our previous theoretical studies such as the generalized Green-Kubo formula~\cite{Chong09b,HO12} are only valid for stationary external forces and is numerically verified for a plane shear, but the most of experiments adopt vibrating granular gases\cite{mennon04,kumar11,joubaud12,mounier12}.
%We also need to demonstrate that the previous formulation can be used even for hard core granular gases.
In this paper, thus, we re-derive the fluctuation relations for soft core granular gases under vibrations around a nonequilibrium steady state.
We also derive the generalized Green-Kubo formula for vibrating beds.

The organization of this paper is as follows.
In Sec. II, we summarize the general framework of Liouville equation and some identities which are used in this paper.
Section III which is the main part of this paper consists of two parts. 
In the first part (Sec. III A) we discuss the the derivation of the integral fluctuation theorem (IFT).
In the second part and the third parts,  we also derive a standard fluctuation theorem (Sec. III B) and the generalized Green-Kubo formula (Sec. III C), respectively.
In Sec. IV we discuss our results and we give conclusion in section V.
In Appendix A, we briefly summarize some operators' identities. %In Appendix B we present the proof of Eq.(\ref{<Omega>=0}).]
%In Appendix C, we discuss the effect of the measurement to get the generalized Jarzynski equality.

\section{Liouville equation}

%\subsection{General framework for the Liouville equation }

Let us consider a system of $N$ identical soft spherical and smooth dissipative particles.
We assume that particles are monodispersed, which are characterized by their diameter $d$ and the mass $m$. 
%and coefficient of restitution $e$ less than unity with its mass $m$.% in $d-$dimensional space.
The particles are influenced by the gravity with the acceleration constant $g$ in $z-$direction
% where the gravitational force is represented by
%$A\omega^2 \cos \omega t$ .
If we use a box fixed frame, each particle feels the acceleration $-g+A\omega^2 \cos\omega t$ in $z-$direction with the amplitude $A$ and the angular acceleration $\omega$.
Moreover, we should introduce a confined potential which prevents particles from penetrating the bottom plate.

The basic equation for the statistical mechanics of frictionless granular particles under such a vibrations is  the Liouville equation.\cite{Evans08,HO12,HO2008,Chong_etal_in_preparation,Suzuki-Hayakawa}  
%The Liouville equation for granular fluidswas investigated by Brey {\it et al.}\cite{brey97,dufty06,dufty06b} some time ago.
The argument in this section is parallel to those in Refs.\cite{Evans08,HO12}.
Let $i\cl (t)$ be the total Liouvillian which operates an arbitrary function $A(\Vect{\Gamma}(t))$ 
starting from $t=0$ as
\begin{equation}\label{Liouville}
\frac{dA(\Vect{\Gamma}(t))}{d{t}}=U_\rightarrow(0,t) i \cl(t) A(\Vect{\Gamma}) , 
\quad 
A(\Vect{\Gamma}(t))=U_\rightarrow(0,t)
%T_\rightarrow e^{i\int_{-\infty}^td\tau \cl_{\rm tot}(\tau)}
 A(\Vect{\Gamma}), 
\end{equation}
where 
\begin{eqnarray}
U_\rightarrow(0,t) &\equiv& T_\rightarrow e^{i \int_{0}^tds \cl(s)} \nonumber\\
&=&
\sum_{n=0}^{\infty}
\int_{0}^t ds_1 \int_{0}^{s_1}ds_2 
\cdots \int_{0}^{s_{n-1}} ds_n 
i{\cal L}(s_n) \cdots i {\cal L}(s_2)i {\cal L}(s_1)
,
\end{eqnarray}
 and
$\Vect{\Gamma}(t)=\{\boldsymbol{r}_i(t),\boldsymbol{p}_i(t)\}_{i=1}^N$ %is the set of the positions $\boldsymbol{r}_i(t)$ and the momenta $\boldsymbol{p}_i(t)$ for $N$ particles 
with the abbreviation $\Vect{\Gamma}\equiv \Vect{\Gamma}(0)$.  
We note that there are some trivial relations for $U_\rightarrow(t_0,t)$ such as
\begin{equation}\label{eq:2.2}
U_\rightarrow(t_0,t)=U_\rightarrow(t_0,s)U_\rightarrow(s,t); \quad
U_\rightarrow(t_0,t) \tilde{ f}(\Vect{\Gamma}(t_0))=\tilde{ f}(\Vect{\Gamma}(t))
\end{equation}
for an arbitrary function $\tilde{ f}(\Vect{\Gamma}(t))$.

The total Liouvillian consists of three parts, the elastic part, the viscous part and the part from an external vibration. 
We write $i\cl(t)$ as 
\begin{equation}
i\cl(t)=i\cl^{\rm (el)}(\bs{\Gamma})+i\cl^{\rm (vis)}(\bs{\Gamma})+i\cl^{({\rm ext})}(\bs{\Gamma},t),
\end{equation}
 where
$i\cl^{\rm (el)}(\bs{\Gamma})$ is   the elastic collision part, 
\begin{equation}\label{cl}
i\cl^{\rm (el)}(\bs{\Gamma})=\sum_{i=1}^N\frac{\bs{p}_i}{m}\cdot\frac{\partial}{\partial \boldsymbol{r}_i}+\bs{F}^{\rm (el)}_i\cdot\frac{\partial}{\partial \bs{p}_i} .
\end{equation}
%where $m$ is the mass of each grain.
Here, we assume that the elastic force can be represented by the summation of the pairwise force
$\bs{F}_i^{\rm (el)}=\sum_{j\ne i}\bs{F}_{ij}^{\rm ( el)}$ with
\begin{equation}
\bs{F}_{ij}^{\rm ( el)}=-\frac{\partial u(r_{ij})}{\partial \bs{r}_{ij}}=\Theta(d-r_{ij})f(d-r_{ij})\hat{\bs{r}}_{ij},
\end{equation}
where we have introduced the pair-wise potential $u(r_{ij})$, %$d$ for the diameter of each grain, 
$\bs{r}_{ij}\equiv \bs{r}_i-\bs{r}_j$, $r_{ij}\equiv |\bs{r}_{ij}|$, $\hat{\bs{r}}_{ij}=\bs{r}_{ij}/r_{ij}$, and
the Heviside function $\Theta(x)$ which satisfies $\Theta(x)=1$ for $x>0$ and $\Theta(x)=0$ for otherwise.
The elastic repulsive force $f(x)$ is proportional to $x$ for the linear spring model, or to $x^{3/2}$ for the Hertzian contact model.

%with the peculiar momentum $\boldsymbol{p}_i$ of the $i-$th particle and the elastic force $\bs{F}_i^{\rm (el)}=\sum_{j\ne i} \bs{F}_{ij}^{\rm (el)}$ acting on $i$-th particle with the force $\bs{F}_{ij}^{\rm (el)}$ acting from $j$-th particle to $i$-th particle. 
Similarly, the viscous Liouvillian $i\cl^{\rm (vis)}$ is the contribution of inelastic collisions:
\begin{equation}\label{viscous-L}
i\cl^{\rm (vis)}(\bs{\Gamma})=\sum_{i=1}^N \bs{F}^{\rm (vis)}_i\cdot\frac{\partial}{\partial \bs{p}_i},
\end{equation}
where  $\bs{F}_i^{\rm (vis)}$ is the viscous force acting on $i-$th particle represented by
$\bs{F}_i^{\rm (vis)}=\sum_{j\ne i}\bs{F}_{ij}^{\rm (vis)}$ with
\begin{eqnarray}\label{F_vis} 
\bs{F}_{ij}^{\rm (vis)}&=&-\hat{\bs{r}}_{ij}\Theta(d-r_{ij})\zeta(d-r_{ij})(\bs{v}_{ij}\cdot\hat{\bs{r}}_{ij}).
\nonumber\\
&=& -\hat{\bs{r}}_{ij}{\cal F}(r_{ij})(\bs{v}_{ij}\cdot\hat{\bs{r}}_{ij}) .
\end{eqnarray}
Here we have introduced $\bs{v}_{ij}\equiv \dot{\bs{r}}_{ij}=d\bs{r}_{ij}/dt$, and
\begin{equation}
{\cal F}(r)\equiv \Theta(d-r)\zeta(d-r)
\end{equation}
with the viscous function $\zeta(x)$ which is a constant or $\zeta(x)\propto x^{1/2}$ corresponding to the linear spring model or the Hertzian contact model for elastic contact force.
The Liouville operator representing the vibration $i\cl^{({\rm ext})}(t)$ is given by
\begin{equation}\label{L_s}
i\cl^{({\rm ext})}(\bs{\Gamma},t)=\sum_{i=1}^N \bs{F}^{\rm (ext)}_i(t)\cdot\frac{\partial}{\partial \bs{p}_i},
%\dgam\sum_{j=1}^N\left(y_j\frac{\partial}{\partial x_j}-p_{y,j}\frac{\partial}{\partial p_{x,j}}\right) ,
\end{equation}
where the vibrating force is given by
\begin{equation}
\bs{F}_i^{\rm (ext)}(t)=\hat{z}\left\{ m (-g+A \omega^2 \cos \omega t) -\frac{\partial V_{\rm ext}(z_i)}{\partial z_i} \right\}
=\hat{z}F_i^{\rm (ext)}(t)
\end{equation}
in a box fixed frame, %with the amplitude $A$, the gravitational acceleration $g$, and the angular frequency $\omega$, 
where $\hat{z}$ is the unit vector in $z$ direction, and $V_{\rm ext}(z)$ represents a confined potential in a box such as
\begin{equation}
V_{\rm ext}(z)=V_0 \exp\left[ -z/\xi \right]
\end{equation}
to prevent grains from penerating the bottom plate of the container.

It should be noted that the Liouvillian is not self-adjoint, 
%in contrast to the case of elastic particles
%, {\it i.e.} the adjoint Liouvillian is different from the Liouvillian 
because of the violation of time-reversal symmetry for each collision. 
The adjoint Liouvillian is defined through the equation of the phase function or the $N-$body distribution function 
$\rho(\Vect{\Gamma},t)$
\begin{equation}\label{rho}
\rho(\Vect{\Gamma},t)=
\tilde{U}_\leftarrow(t,0)\rho(\Vect{\Gamma},0), \qquad 
\frac{\partial \rho(\Vect{\Gamma},t)}{\partial t}=-i{\cl}^\dagger(t) \rho(\Vect{\Gamma},t),
\end{equation}
where 
\begin{eqnarray}
\tilde{U}_\leftarrow(t,0)&=&T_\leftarrow e^{-i \int_{0}^tds \cl^\dagger(s)} \nonumber\\
 &\equiv&
\sum_{n=0}^{0}(-)^n\int_{0}^t ds_1\int_{0}^{s_1}ds_2\cdots \int_{0}^{s_{n-1}}ds_n i\cl^\dagger(s_1) i\cl^\dagger(s_2)\cdots i \cl^\dagger(s_n) .
\end{eqnarray}

The adjoint Liouvillian satisfies
\begin{equation}\label{eq:17}
i\cl^\dagger(\bs{\Gamma},t)=i{\cl}(\bs{\Gamma},t)+\Lambda(\Vect{\Gamma}) ,
\end{equation}
where 
\begin{equation}\label{Lambda}
\Lambda(\Vect{\Gamma})\equiv \frac{\partial}{\partial \bs{\Gamma}} \cdot \dot{\bs{\Gamma}}(\bs{\Gamma})
\end{equation} 
is the phase volume contraction. 
We note that  $\Lambda(\bs{\Gamma})$ in our system 
does not depend on $t$ explicitly, which can be written as
\begin{equation}
\Lambda(\bs{\Gamma}) = \sum_{i} \frac{\partial}{\partial \bs{p}_{i}} \cdot \bs{F}_{i}^{\rm (vis)} =
- \frac{1}{m} \sum_{i} \sum_{j \ne i} {\cal F}(r_{ij})
\label{eq:SLLOD-Lambda}
\end{equation}
for $t\ge 0$.
The phase volume contraction $\Lambda(\bs{\Gamma})$ is directly related to the change of Jacobian
\begin{equation}\label{Jacobian}
\left|\frac{\partial \bs{\Gamma}(t)}{\partial \bs{\Gamma}}\right|
= \exp\left[\int_0^td\tau \Lambda(\bs{\Gamma}(\tau)) \right],
\end{equation}
where $\Lambda(\bs{\Gamma}(t))=U_\rightarrow(0,t)\Lambda(\bs{\Gamma})U_\leftarrow(t,0)$.
Note that the time evolution of an arbitrary physical function $A(\bs{\Gamma}(t))$ is given by 
 $A(\bs{\Gamma}(t))=U_\rightarrow(0,t)A(\bs{\Gamma})U_\leftarrow(t,0)$,
 where we have introduced $U_\leftarrow(t,0)\equiv T_\leftarrow \exp[-i\int_0^tds \cl (s)]=U_\rightarrow^{-1}(0,t)$.

%where $\rho(\Gamma(t))$ is the (real) phase function which is $N-$body distribution function.
The average of a physical quantity is defined as
\begin{equation}\label{average}
\langle A(\bs{\Gamma}(t))\rangle \equiv \int d\Vect{\Gamma} \rho(\Vect{\Gamma},0)A(\Vect{\Gamma}(t))=\int d\Vect{\Gamma} A(\Vect{\Gamma})\rho(\Vect{\Gamma},t) .
\end{equation}
From Eqs. (\ref{Liouville}), (\ref{rho}) and (\ref{average}) we obtain the relations
\begin{equation}\label{average2}
\int d\Vect{\Gamma} \rho(\Vect{\Gamma})U_\rightarrow(0,t)A(\Vect{\Gamma})=\int d\Vect{\Gamma} A(\Vect{\Gamma})
\tilde{U}_\leftarrow(t,0)\rho(\Vect{\Gamma},0)
\end{equation}
and
\begin{equation}\label{average3}
\int d\Vect{\Gamma} \rho(\Vect{\Gamma})i\cl(t) A(\Vect{\Gamma})=-\int d\Vect{\Gamma} A(\Vect{\Gamma}) i\cl^\dagger(t) \rho(\Vect{\Gamma},0) .
\end{equation}
%Note that Eq. (\ref{average}) can be used for any $t$.
% and does not have to assume the starting point from $t=0$ as in Eq.(\ref{average2})by replacing $\bs{\Gamma}$ by $\bs{\Gamma}(t_0)$. 
%We also have the following relation for the inverse path:
%\begin{equation}\label{claasical-13-b}
%\langle \check{A}(\bs{\Gamma}(-t)) \rangle
%=\int d\mathbf{\Gamma} \rho(\mathbf{\Gamma},0) \{ U_{\leftarrow}(t,0)A(\mathbf{\Gamma}) \}
%=\int d\mathbf{\Gamma}  A(\mathbf{\Gamma}) \{ \tilde{U}_\rightarrow (0,t)\rho(\mathbf{\Gamma},0)\} ,
%\end{equation}
%where $\check{A}(\bs{\Gamma}(-t))=U_\leftarrow(t,0)A(\bs{\Gamma})=U_\leftarrow(t,0)A(\bs{\Gamma})U_\rightarrow(0,t)$
%and $\check{A}(\bs{\Gamma}(t))=A(\bs{\Gamma}(t))=U_\rightarrow(0,t)A(\bs{\Gamma})U_\leftarrow(t,0)$ for $t\ge 0$.
%Note that $\check{A}(\bs{\Gamma}(-t))$ is not equal to $A(\bs{\Gamma}(-t))$ except for the case that the system has time translational symmetry.
%Therefore,  
%$U_\leftarrow(0,-t)=U_\rightarrow^{-1}(-t,0)$ and $\tilde{U}_\rightarrow(-t,0)=\tilde{U}_\leftarrow^{-1}(0,-t)$ are, respectively, not equal to
%$U_\leftarrow(t,0)=U_\rightarrow^{-1}(0,t)$ and $\tilde{U}_\rightarrow(0,t)=\tilde{U}_\leftarrow^{-1}(t,0)$ in general,
%where $\tilde{U}_\rightarrow(0,t)\equiv T_\rightarrow \exp[i\int_0^t ds \cl^\dagger(s) ]$.
%In other words, the useful relations $U_\leftarrow(t,0)=U_\leftarrow(0,-t)$ and $\tilde{U}_\rightarrow(-t,0)=\tilde{U}_\leftarrow(0,t)$ can be used, 
%only if the Liouville operator is independent of time.
%However, such relations cannot be held in general situations. 

Let us introduce a stationary distribution to characterize the quasi-periodic motion of granular particles under the periodic vibration.
In the stationary process, the initial distribution function $\rho(\bs{\Gamma},0)$ may have the form
\begin{equation}
\rho(\Vect{\Gamma},0)=\rho_{\rm ini}(\Vect{\Gamma})\equiv \frac{e^{-  I_0(\mathbf{\Gamma})}}{\cal Z} ,
\label{large-deviation}
\end{equation} 
where $I_0(\bs{\Gamma})\equiv I(\bs{\Gamma},t=2n\pi/\omega)$ with an arbitrary integer $n$ and am effective potential $I(\bs{\Gamma},t)$.
% and
%\begin{equation}
%I(\Vect{\Gamma},t)\equiv %\sum_i \frac{\bs{p}_i^2}{2m}+\frac{1}{2}\sum_{i,j}u(r_{ij})
%I_0(\bs{\Gamma})+\sum_i \int_{\tau=0}^{\tau=t}  d\bs{r}_i\cdot\bs{F}_i^{\rm (ext)}(\tau)
%\end{equation}
% is the effective potential, 
and ${\cal Z} \equiv \int d\Vect{\Gamma}e^{-I_0(\mathbf{\Gamma})}$.
% with ${\cal Z}\equiv {\cal Z}(t=0)$.
Note that $I_0(\bs{\Gamma})$ is an arbitrary function of $\bs{\Gamma}$, and thus, this choice is quite general for the argument.
In the stationary process, we assume that an average of an arbitray function $A(\bs{\Gamma}(t))$ satisfies the perodic condition
\begin{equation}\label{periodic_average}
\left\langle A
\left(
\bs{\Gamma}
\left(
t+\frac{2n\pi}{\omega} 
\right) 
\right)
\right\rangle =\langle A(\bs{\Gamma}(t) \rangle
\end{equation}
for any nonzero integer $n$. 
This assumption is reasonable because of the periodicity of the Liouvillian:  
\begin{equation}\label{periodic_L}
i\cl\left(t+\frac{2n\pi}{\omega}\right)=i\cl(t)
\end{equation}
for an arbitrary integer $n$.
We should note that $U_\leftarrow(0,-t)=U_\rightarrow^{-1}(-t,0)$ and $\tilde{U}_\rightarrow(-t,0)=\tilde{U}_\leftarrow^{-1}(0,-t)$ are, respectively, not equal to
$U_\leftarrow(t,0)=U_\rightarrow^{-1}(0,t)$ and $\tilde{U}_\rightarrow(0,t)=\tilde{U}_\leftarrow^{-1}(t,0)$ in general,
where  $U_\leftarrow(t,0)\equiv T_\leftarrow e^{-i\int_0^t d\tau \cl (\tau)}$ and  $\tilde{U}_\rightarrow(0,t)\equiv T_\rightarrow \exp[i\int_0^t ds \cl^\dagger(s) ]$.
However, we can use 
\begin{equation}\label{stationary-inverse}
U_\leftarrow(t,0)=U_\leftarrow(0,-t) \quad {\rm and} \quad
 \tilde{U}_\rightarrow(-t,0)=\tilde{U}_\rightarrow(0,t)
\end{equation}
 for stationary state for $t=2n\pi/\omega$ with an integer $n$.
 Indeed, with the aid of Eq.(\ref{periodic_L}) we can rewrite
$U_\leftarrow(t,0)
 =T_\leftarrow e^{-i \int_{-2n\pi/\omega}^{t-2n\pi/\omega}d\tau' \cl(\tau')}$.
 Furthermore, if we restrict the time for the measurement to $t=2n\pi/\omega$, we can rewrite 
$U_\leftarrow(t,0)=T_\leftarrow e^{-i\int_{-t}^0d\tau \cl(\tau)}=U_\leftarrow(0,-t)$.

In the last part of this subsection, we introduce a useful formula between $\tilde{U}_\rightarrow(0,t)$ and $U_\rightarrow(0,t)$.
Any two Liouville operators even for $i\cl(t)$ and $i\cl^\dagger(t)$ satisfy Dyson's equation\cite{Evans08}:
\begin{equation}\label{dyson}
\tilde{U}_\rightarrow(0,\tau)=U_\rightarrow(0,\tau)+\int_{0}^{\tau}ds \tilde{U}_\rightarrow(0,s)
\Lambda(\Vect{\Gamma})
U_\rightarrow(s,\tau) ,
\end{equation}
where we have used Eq.(\ref{eq:17}). It is straightforward to rewrite Eq.(\ref{dyson}) as\cite{Evans08}
\begin{equation}\label{kawasaki}
\tilde{U}_\rightarrow(0,t)=\exp
\left[\int_0^td\tau \Lambda(\bs{\Gamma}(\tau))  \right]
U_\rightarrow(0,t) .
\end{equation}
where $\Lambda(\bs{\Gamma}(t))=U_\rightarrow(0,t)\Lambda(\bs{\Gamma})U_\leftarrow(t,0)$.

\section{Fluctuation Theorem and Green-Kubo formula}

Now, let us derive some important identies such as fluctuation relations and the generalized Green-Kubo formula for vibrating granular materials.
%Fluctuation theorem can be derived from Eq.(\ref{claasical-13-b}) coupled with Eq. (\ref{kawasaki}).
For the demonstration of the existence of the above mentioned identities, we explain the derivations as follows. 
The first part is dedicated to the derivation of the integral fluctuation theorem (IFT).
%In the second part, we derive a formal steady distribution and its explicit expression in terms of the perturbation method.
We also derive both the standard fluctuation theorem in the second part and the generalized Green-Kubo formula in the third part.

\subsection{Integral fluctuation theorem}

%THIS SECTION IS STILL UNCLEAR. IN WHAT SITUATION WE NEED THE TIME DEPENDENCE OF THE NORMALIZATION?

The integral fluctuation theorem (IFT) is one of representations of the fluctuation theorem, which is directly related to 
Jarzynski equality.\cite{Jarzynski}
The relation between Jarzynski equality and the fluctuation theorem has been investigated extensively.\cite{Crooks,Seifert12}
Although IFT is an important identity for the stationary sheared granular systems\cite{Chong09b}, and
this equality plays a fundamental role even in granular systems under a vibration.
% though we do not present the details of it. 

To demonstrate the existence of the IFT, we consider a system characterized by the following time-dependent Hamilitonian :
\begin{equation}\label{t-Hamilitonian}
H_0(\bs{\Gamma}(t))=\sum_i \frac{{\bs{p}_i(t)}^2}{2m}+\frac{1}{2}\sum_{i,j\ne i}u(r_{ij}(t)).
\end{equation}
We also assume that the initial condition satisfies the canonical distribution 
\begin{equation}\label{canonical}
\rho_{\rm eq}(\bs{\Gamma})=\frac{e^{-\beta H_0(\bs{\Gamma})}}{Z(\beta)}
,
\end{equation}
where $\beta$ is the inverse temperature and $Z(\beta)\equiv \int d\bs{\Gamma}e^{-\beta H_0(\bs{\Gamma})}$.
%We note that the phase volume contraction (\ref{eq:SLLOD-Lambda}) and the Jacobian (\ref{Jacobian}) are unchanged in this argument, 
%because we only modify the Hamitonian and the dissipative part is unchanged.

In this case, it is easy to verify the conservation of the normalization factor. i.e. 
$Z(\beta)=\int d\bs{\Gamma} e^{-\beta H_0(\bs{\Gamma})}=\int d\bs{\Gamma}(t) e^{-\beta H_0(\bs{\Gamma}(t)}$.
Because the time derivative of $H_0(\bs{\Gamma}(t))$ is given by
\begin{equation}
\dot{H}_0(\bs{\Gamma}(t))= \sum_i v_{i,z}(t)F_i^{(\rm ext)}(t)-2 {\cal R}(\bs{\Gamma}(t))
\end{equation}
with $\dot{H}_0\equiv dH_0/dt$, $\bs{v}_i(t) \equiv d\bs{r}_i(t)/dt$, 
\begin{equation}\label{Rayleigh}
{\cal R}(\bs{\Gamma})\equiv 
-\frac{1}{4}\sum_{i,j}\bs{v}_{ij}\cdot\bs{F}_{ij}^{({\rm vis})}
=\frac{1}{4}\sum_{i,j} {\cal F}(r_{ij}) (\bs{v}_{ij} \cdot\hat{\bs{r}}_{ij})^2 ,
\end{equation} 
and $\bs{v}_{ij}\equiv \bs{v}_i-\bs{v}_j$,
the conservation of the probability leads to
\begin{eqnarray}\label{Jarzynski_derivation}
1&=&\int d\bs{\Gamma}(t) \frac{e^{-\beta H_0(\bs{\Gamma}(t))}}{Z(\beta)}
\nonumber\\
&=&
\int d\bs{\Gamma}\left|\frac{\partial{\bs{\Gamma}(t)}}{\partial \bs{\Gamma}} \right|
\frac{e^{-\beta H_0(\bs{\Gamma})}}{Z(\beta)}
%\nonumber\\
%&& \times
\exp\left[ \beta \int_0^t d\tau \{ \sum_i v_{i,z}(\tau)F_i^{(\rm ext)}(\tau)-2 {\cal R}(\bs{\Gamma}(\tau)) \} \right]
\nonumber\\
&=& 
\int d\bs{\Gamma} \frac{e^{-\beta H_0(\bs{\Gamma})}}{Z(\beta)}\exp\left[
-\int_0^t d\tau \Omega_{\rm eq}(\bs{\Gamma}(\tau))\right]
\nonumber\\
&=&
\left\langle \exp\left[-\int_0^t d\tau \Omega_{\rm eq}(\bs{\Gamma}(\tau)) \right] \right\rangle_{\rm eq} ,
\end{eqnarray} 
where we have used Eq.(\ref{Jacobian}) for the third equality, and introduced
\begin{equation}\label{Omega_bar}
\Omega_{\rm eq}(\bs{\Gamma}(t)) 
\equiv 
 \beta \sum_i v_{i,z}(t)F_i^{(\rm ext)}(t)-2\beta {\cal R}(\bs{\Gamma}(t))-\Lambda(\bs{\Gamma}(t)) ,
\end{equation} 
and the average $\langle \cdot \rangle_{\rm eq}\equiv \frac{1}{Z(\beta)} \int d\bs{\Gamma} e^{-\beta H_0(\bs{\Gamma})} \cdot .$ 
Note that this derivation differs from those presented in Ref.\cite{HO12}.
%with

Equation (\ref{Jarzynski_derivation}) associated with Eq.(\ref{Omega_bar}) is the IFT for granular fluids under the vibration.
%We should note that $\bs{\Gamma}(t)$ in this subsection is equivalent to $\bs{\Gamma}_+(t)$ because the evolution dynamics for $t>0$ is governed by $i\cl_{\rm st}$.
%We also note that the time dependent term $\sum_i g(t) z_i(t)$ in ${\cal H}(t)$ is necessary to change $Z(\beta;t)$.
%Namely, $Z(\beta;t)=Z(\beta;0)$ if $g(t)=0$, which is reduced to the IFT. 
It is a characteristic feature for dissipative systems that the phase volume contraction $\Lambda(\bs{\Gamma})$ is involved in $\Omega_{\rm eq}(\bs{\Gamma}(t))$. Thus, the right hand side of the IFT (\ref{Jarzynski_derivation}) for dissipative cases cannot be represented by the work done by the external force.

The IFT (\ref{Jarzynski_derivation}) is directly reduced to an inequality  
\begin{equation}\label{2ndlaw}
\int_0^t d\tau \langle \Omega_{\rm eq}(\bs{\Gamma}(\tau)) \rangle \ge 0
\end{equation}
with the aid of Jenssen's inequality.
This inequality ensures the existence of an entropy-like quantity even granular systems under the vibration.

Now, let us extend the IFT to the case of starting from an arbitrary distribution $\rho_{\rm ini}(\bs{\Gamma})$.
In this case, the IFT can be rewritten as
\begin{eqnarray}\label{Jarzynski_SS}
1&=&
%\frac{{\cal Z}(t)}{{\cal Z}(0)}=
\int d\bs{\Gamma}(t) \frac{e^{-I_0(\bs{\Gamma}(t))}}{{\cal Z}}
\nonumber\\
&=&
\int d\bs{\Gamma}\left|\frac{\partial{\bs{\Gamma}(t)}}{\partial \bs{\Gamma}} \right|
\frac{e^{-I_0(\bs{\Gamma})}}{{\cal Z}}\exp\left[ -\int_0^t d\tau \dot{I}_0(\bs{\Gamma}(\tau)) \right]
\nonumber\\
&=& 
\int d\bs{\Gamma}\frac{e^{-I_0(\bs{\Gamma})}}{{\cal Z}}\exp\left[
-\int_0^t d\tau {\Omega}(\bs{\Gamma}(\tau))\right]
\nonumber\\
&=&
\left\langle \exp\left[-\int_0^t d\tau {\Omega}(\bs{\Gamma}(\tau)) \right] \right\rangle ,
\end{eqnarray} 
where 
$\Omega(\bs{\Gamma}(t))=\dot{I}_0(\bs{\Gamma}(t))-\Lambda(\bs{\Gamma}(t))$, 
 and ${\cal Z}\equiv \int d\bs{\Gamma} e^{-I_0(\bs{\Gamma})}=\int d\bs{\Gamma}(t) e^{-I_0(\bs{\Gamma}(t))}$
and $\langle \cdot \rangle =\int d\bs{\Gamma} \frac{e^{-I_0(\bs{\Gamma})}}{{\cal Z}} \cdot$. 
%Thus, it is easy to extend the Jarzynski equality starting from an arbitrary distribution.
Equation (\ref{Jarzynski_SS}) is also reduced to the entropy-like relation
\begin{equation}
\int_0^t d\tau \langle \Omega(\bs{\Gamma}(\tau)) \rangle \ge  0 .
\end{equation}

\subsection{The standard Fluctuation Theorem }

The direct consequences of Eq. (\ref{Jarzynski_SS}) are two important relations, the conventional fluctuation theorem and the generalized Green-Kubo formula, from Eq.(\ref{Jarzynski_SS}). Here, let us illustrate how to derive these relations.

It is straightforward to derive the conventional fluctuation theorem (FT) from IFT (\ref{Jarzynski_derivation}) or (\ref{Jarzynski_SS}), where FT is the relation of the probability of the entropy production  between the forward path and the inverse path \cite{Evans08}.
Because the derivation of FT starting from a canonical distribution has already been discussed in Ref.\cite{HO12}, 
we, here,  present the derivation of FT from Eq.(\ref{large-deviation}) under the assumption that $\rho_{\rm ini}(\bs{\Gamma})$ is invariant by the time reversal operation.
Of course, the outline of the derivation is unchanged.

Now let us consider the process from time 0 to time $t$ by the time evolution operator $U_\rightarrow(0,t)$ and the trajectory of the phase variable $\bs{\Gamma}(\tau)=U_\rightarrow(0,t)\bs{\Gamma}$ for $0\le \tau \le t$.
The inverse process, thus, is characterized by the time evolution operator $U_\leftarrow(t,0)$ and the 
 inverse phase variable 
$\bs{\Gamma}^*(\tau)\equiv \{ \bs{r}_i(t-\tau), -\bs{p}_i(t-\tau) \}_{i=1}^N=\{\bs{\Gamma}(t-\tau)\}^T$ for $0\le \tau \le t$,
 where the operation $\{\bs{\Gamma}(t) \}^T$ represents the change of the sign of the momenta
 $\{\bs{\Gamma}(t) \}^T\equiv \{\bs{r}_i(t), -\bs{p}_i(t) \}_{i=1}^N$. 
Because the probability of the inverse trajectories $\rho_{\rm ini}(\bs{\Gamma}^*)
$
 is still normalized as $\int d\bs{\Gamma}^* \rho_{\rm ini}(\bs{\Gamma}^*)=1$ with the abbreviation $\bs{\Gamma}^*\equiv \bs{\Gamma}^*(0)$, 
Eq. (\ref{Jarzynski_SS}) can be rewritten as
\begin{equation}
\int d\bs{\Gamma} \rho_{\rm ini}(\bs{\Gamma}) e^{-t \overline{\Omega_t}}
=\int d\bs{\Gamma}^* \rho_{\rm ini}(\bs{\Gamma}^*) ,
\label{FT_steady2}
\end{equation}
where we have introduced $\overline{\Omega_t}\equiv \frac{1}{t}\int_0^t d\tau \Omega(\bs{\Gamma}(\tau))$
with $ \Omega(\bs{\Gamma}(t))= \dot{I}_0(\bs{\Gamma}(t))-\Lambda(\bs{\Gamma}(t)) $.
%With the aid of the time reversal symmetry for $\rho_{\rm eq}(\bs{\Gamma})=\rho_{\rm eq}(\bs{\Gamma}^T)$ and $d\bs{\Gamma}^T=d\bs{\Gamma}|\partial \bs{\Gamma}^T/\partial \bs{\Gamma}|$, 
%Eq. (\ref{FT_steady2}) means that the Jacobian $|\partial \bs{\Gamma}^T/\partial \bs{\Gamma}|$ is given by $\exp[-t\overline{\Omega}_t]$.
%Thus, the integrand of Eq.(\ref{FT_steady2}) satisfies $d\bs{\Gamma} \rho_{\rm eq}(\bs{\Gamma}) e^{-t \overline{\Omega_t}}
%=d\bs{\Gamma}^T \rho_{\rm eq}(\bs{\Gamma}^T)$. 
From the definition of $\Lambda(\bs{\Gamma}(t))$ and the assumption $I_0(\bs{\Gamma}^*(\tau))=I_0(\bs{\Gamma}(t-\tau))$
, there are some trivial relations: 
$\Lambda(\bs{\Gamma}^*(\tau))=-\Lambda(\bs{\Gamma}(t-\tau))$,
 and
$\Omega(\bs{\Gamma}^*(\tau))=-\Omega(\bs{\Gamma}(t-\tau))$ for $0\le \tau \le t$. 
%Here, $\bs{\Gamma}_+(\pm t)=e^{\pm i\cl_+t}\bs{\Gamma}$.
Therefore, we can write the probability of $\overline{\tilde{\Omega}_t}=-A$ for 
%$\tilde{\Omega}_{\rm eq}(\bs{\Gamma})\equiv \Omega_{\rm eq}(\bs{\Gamma}^T)$ and 
$\overline{\tilde{\Omega}_t}\equiv \frac{1}{t}\int_0^td\tau {\Omega}(\bs{\Gamma}^*(\tau))
=\frac{1}{t}\int_0^t d\tau \Omega(\{\bs{\Gamma}(\tau)\}^T) $: 
\begin{eqnarray}\label{conventional_FT}
{\rm Prob}(\overline{\tilde{\Omega}_t}=-A)
&=&\int d\bs{\Gamma}^* \rho_{\rm ini}(\bs{\Gamma}^*)\delta(\overline{\tilde{\Omega}_t}+A)
\nonumber\\
&=& \int d\bs{\Gamma}(t) \rho_{\rm ini}(\bs{\Gamma}(t))\delta(\overline{{\Omega}_t}-A) \nonumber\\
&=& \int d\bs{\Gamma} \rho_{\rm ini}(\bs{\Gamma})e^{-t\overline{\Omega}_t}\delta(\overline{\Omega_t}-A)
\nonumber\\
&=& e^{-At}\int d\bs{\Gamma}\rho_{\rm ini}(\bs{\Gamma})\delta(\overline{\Omega_t}-A)
\nonumber\\
&=& e^{-At} {\rm Prob}(\overline{\Omega_t}=A) 
\end{eqnarray}
for the conventional fluctuation theorem, where we have used $|\partial \bs{\Gamma}^*(\tau)/\partial \bs{\Gamma}(t-\tau)|=1$,
$\rho_{\rm ini}(\bs{\Gamma}(t))=\rho_{\rm ini}(\bs{\Gamma})e^{-\beta \int_0^t d\tau \dot{H}(\tau)}$
and $ d\bs{\Gamma}(t)= d\bs{\Gamma}e^{\int_0^t d\tau \Lambda(\bs{\Gamma}(\tau))}$.
Note that  the argument is still valid for the general starting point Eq. (\ref{large-deviation}) if we have the symmetry $I(\bs{\Gamma}^*(\tau))=I(\bs{\Gamma}(t-\tau))$. 

%Note that our fluctuation theorem 
%\begin{equation}\label{form_FT}
%{\rm Prob}(\overline{\tilde{\Omega}_t}=A)=e^{-At} {\rm Prob}(\overline{\Omega_t}=A) 
%\end{equation}
%differs from the standard form ${\rm Prob}(\overline{\tilde{\Omega}_t}=-A)=e^{-At}{\rm Prob}(\overline{\Omega_t}=A)$, because our system does not have the time reversal symmetry.

\subsection{Generalized Green-Kubo formula}

Next, let us derive the generalized Green-Kubo formula from Eq.(\ref{Jarzynski_SS}) following the argument in Ref.\cite{Chong09b,HO12}.
It should be noted that the generalized Green-Kubo formula is only valid for $t=2n\pi/\omega$ with an arbitrary integer $n$, i.e. at time with an identical phase of oscillation.
Nevertheless, the argument in this subsection can be used for time dependent processes which has not be proven in  Ref.\cite{HO12}.

Now, let us rewrite Eq.(\ref{rho}) as
\begin{eqnarray}\label{GK_1}
\rho_{\rm ini}(\bs{\Gamma})
&=&
\tilde{U}_\rightarrow(0,t)\rho(\bs{\Gamma},t)
= e^{\int_0^td\tau \Lambda(\bs{\Gamma}(\tau))}\rho(\bs{\Gamma}(t),t) ,
\end{eqnarray}
where we have used the identity (\ref{kawasaki}).
Let us operate $U_\leftarrow(t,0)$ on the both side of Eq.(\ref{GK_1}) with the aid of Eq.(\ref{stationary-inverse}) at $t=2n \pi/\omega$ with an integer $n$, we can write
\begin{eqnarray}\label{U_rho}
U_\leftarrow(t,0)\rho_{\rm ini}(\bs{\Gamma})&=& U_\leftarrow(0,-t)\rho_{\rm ini}(\bs{\Gamma})=\rho_{\rm ini}(\bs{\Gamma}(-t))
\nonumber\\
&=& \frac{e^{-I_0(\bs{\Gamma}(-t))}}{{\cal Z}}= 
e^{\int_0^t d\tau \dot{I}_0(\bs{\Gamma}(-\tau))}\rho_{\rm ini}(\bs{\Gamma})
\nonumber\\
&=& U_\leftarrow(t,0)[e^{\int_0^t d\tau \Lambda(\bs{\Gamma}(t-\tau))}\rho(\bs{\Gamma}(t),t)]
=e^{\int_0^t d\tau \Lambda(\bs{\Gamma}(-\tau))}\rho(\bs{\Gamma},t)
\end{eqnarray}
From Eq.(\ref{U_rho}) we immediately obtain
\begin{equation}\label{rho_time}
\rho(\bs{\Gamma},t)=e^{\int_0^t d\tau \Omega(\bs{\Gamma}(-\tau))}\rho_{\rm ini}(\bs{\Gamma}) . 
\end{equation}
%The expression (\ref{rho_time}) leads to the conservation of the probability:
%\begin{equation}\label{conservation_prob}
% \left\langle \exp\left[\int_0^td\tau {\Omega}(\bs{\Gamma}(-\tau)) \right] \right\rangle=1 .
%\end{equation}

%The generalized Green-Kubo formula is the direct consequence of the IFT.
%For simplicity, we restrict our interest to the case that the external force is conservative. In this case
%Because the derivation has already been discussed for the classical case\cite{chong}, here we demonstrate
%how to derive the generalized Green-Kubo formula for the quantum case. 
The differentiation of Eq.(\ref{average}) with the help of Eq.(\ref{rho_time}), 
we obtain
\begin{eqnarray}
\frac{d}{dt}\langle A(\bs{\Gamma}(t)) \rangle
&=&
\int d\Vect{\Gamma} %\left(
A(\Vect{\Gamma}){\Omega}(\Vect{\Gamma}(-t)) \rho(\Vect{\Gamma},t) %\right)
=
\int d\bs{\Gamma} U_\leftarrow(t,0)\{A(\bs{\Gamma}(t))\Omega(\bs{\Gamma})\}\rho(\bs{\Gamma},t)
\nonumber\\
&=&
\int d\bs{\Gamma} A(\bs{\Gamma}(t))\Omega(\bs{\Gamma})\tilde{U}_\rightarrow(0,t)\rho(\bs{\Gamma},t)
= \int d\bs{\Gamma} A(\bs{\Gamma}(t))\Omega(\bs{\Gamma}) e^{\int_0^td\tau\Lambda(\bs{\Gamma}(\tau))}\rho(\bs{\Gamma}(t),t)
\nonumber\\
&=& \int d\bs{\Gamma} A(\bs{\Gamma}(t))\Omega(\bs{\Gamma}) \frac{e^{-I_0(\bs{\Gamma})}}{\cal Z} 
=\langle A(\Vect{\Gamma}(t))\Omega(\Vect{\Gamma}) \rangle .
\end{eqnarray}
%where we have used the parallel calculation to that in Eq.(\ref{dY/dt=0}).
This equation can be integrated over $t$ as
\begin{equation}\label{Green-Kubo1}
\langle A(\Vect{\Gamma}(t)) \rangle=\langle A(\Vect{\Gamma}) \rangle
+\int_{0}^tds \langle A(\Vect{\Gamma}(s))\Omega(\Vect{\Gamma}) \rangle .
\end{equation}
This result depends on $\rho_{\rm ini}(\bs{\Gamma})$.
Therefore, the formal response theory can be written as
\begin{equation}\label{Green-Kubo_steady}
\langle \delta A(\bs{\Gamma}) \rangle
=\int_0^\infty dt 
\langle A(\Vect{\Gamma}(t))\Omega(\Vect{\Gamma}) \rangle ,
\end{equation}
where $\delta A(\bs{\Gamma})\equiv \lim_{t\to \infty}A(\bs{\Gamma}(t))-A(\bs{\Gamma})$.

%Thus, the gereneralized Green-Kubo formula should be written as 
%\begin{equation}
%\langle A \rangle_{\rm SS} =\langle A(\Vect{\Gamma}) \rangle_{\rm SS}+\int_{0}^{\infty}ds \langle A(\tau) \Omega(0) \rangle_{\rm SS}
%\label{general-GC}
%\end{equation}
%where $\langle A \rangle_{\rm SS}\equiv \lim_{t\to \infty} \langle A(t) \rangle_{\rm SS}$.
\section{Discussion}

In this paper, we obtain exact nonequilibrium relations.
To verify the validity, we may need numerical simulations as in Ref.\cite{HO12}.
In simulations, we need to restrict our interest to the statistics for small number of particles with large number of sample averages.
For example, Ref.\cite{HO12} use $18$ grains with 800,000 samples.
It should be noted that the verification of the generalized Green-Kubo formula is not difficult by the direct simulation,
but the confirmation of the integral fluctuation theorem by simulations is not easy because of the limitation of numerical accuracy.
%Indeed, the normalization condition (\ref{conservation_prob}) could not be achieved in the simulation even for thermostat systems.\cite{Evans90} 
%There exists no numerical verifications even for the generalized Green-Kubo formula starting not from $\rho_{\rm ini}(\bs{\Gamma})$.
We should stress that these identities can be used even for dense granular systems above the jamming transition.

Although the results obtained in this paper is exact without the limitation of applicability range, 
the actual confirmation for large systems is almost impossible, because the theory requires all cumulants and information for $N$-body distribution function, which are not correctly measured in experiments.
Similarly, to prepare  the general initial distribution function (\ref{large-deviation}) except for the equilibrium condition (\ref{canonical}) experimentally is almost impossible.
In Ref.\cite{HO12}, we use the inverse trajectory in which the time flows from the future to the past, which cannot be used in experiments. 
These difficulties come from the fact that the derived equation is exact without using any coarsening procedure.

Thus, 
it is not easy to calculate the correlation function Eq.(\ref{Green-Kubo1}) or Eq.(\ref{Green-Kubo_steady}).
One of possible methods is to use the mode-coupling theory (MCT).
It is helpful to apply MCT for granular liquids to characterize theology near the jamming transition.\cite{HO2008,Kranz10,Kranz12,SH13,SH13new}   
It is notable that Ref.\cite{Suzuki-Hayakawa} develops a linear response theory for a sheared thermostat system around a nonequilibrium steady state.
The application of this method will be discussed elsewhere.

\section{Summary}

We have developed some exact relations  for frictionless granular fluids under vibrations.
We derived the integral fluctuation theorem and the standard fluctuation theorem around a nonequilibrium steady state.
%We gave a simple derivation of the Jarzynski equality if the Hamiltonian depends on time explicitly.
We finally obtained the generalized Green-Kubo formula around a nonequilibrium steady state.

In this paper, we focus on the detailed analytic calculation on the granular fluids under the vibration.
%The validity of these expressionsss has already been confirmed by numerical simulations in part.
The systematic check in terms of the simulations will be reported elsewhere.

\label{sec:summary}

\begin{acknowledgments}
The author thanks S.-H. Chong, M. Otsuki, K. Suzuki and K. Saitoh for fruitful discussions. 
This work was supported by 
the Grant-in-Aid of MEXT (Grant Nos. 25287098)
and in part by the Yukawa International Program for 
Quark-Hadron Sciences (YIPQS).

%The numerical computations were carried out in computers of YITP, Kyoto University.
\end{acknowledgments}

\appendix

\section{Some operators' identities}

Let us consider the time evolution of $\bs{\Gamma}(t;t_0)$ defined by
\begin{equation}\label{a1}
\bs{\Gamma}(t;t_0)=U_{\rightarrow}(t_0,t)\bs{\Gamma}(t_0) ,
\end{equation}
where we have explicitly written the initial time $t_0$.

By using $U_\rightarrow(t_0,t)$ and $U_\leftarrow(t,t_0)$ it is notable that there is an important relation for $U_\rightarrow(t_0,t)$:
\begin{equation}\label{a_UAU=UA}
A(\bs{\Gamma}(t))\equiv U_\rightarrow(t_0,t)A(\bs{\Gamma}(t_0))U_\leftarrow(t,t_0) = U_\rightarrow(t_0,t)A(\bs{\Gamma}(t_0)) .
\end{equation} 
The proof of (\ref{a_UAU=UA}) is straightforward.
The right hand side of Eq.(\ref{a_UAU=UA}) can be rewritten as
\begin{eqnarray}
U_{\rightarrow}(t_0,t)A(\bs{\Gamma}(t_0))
&=&
U_{\rightarrow}(t_0,t) A(\bs{\Gamma}(t_0) )
  U_{\leftarrow}(t,t_0)U_\rightarrow(t,t_0) 1
=U_\rightarrow(t_0,t)A(\bs{\Gamma}(t_0)) U_\leftarrow(t,t_0) 1
\nonumber\\
&=& U_\rightarrow(t_0,t)A(\bs{\Gamma}(t_0))U_\leftarrow(t,t_0) ,
\end{eqnarray}
where we have used $U_\rightarrow(t_0,t)1=1$ for a constatnt 1.
When we use Eq.(\ref{a_UAU=UA}), we readily obtain
\begin{equation}\label{a_UAB=UA UB}
U_\rightarrow(t_0,t)A(\bs{\Gamma}(t_0)) B(\bs{\Gamma}(t_0))
=A(\bs{\Gamma}(t;t_0)) \cdot B(\bs{\Gamma}(t;t_0)).
\end{equation}
Indeed, the left hand side of this equation can be rewritten as
\begin{eqnarray}
U_\rightarrow(t_0,t)A(\bs{\Gamma}(t_0)) B(\bs{\Gamma}(t_0))
&=& U_\rightarrow(t_0,t) A(\bs{\Gamma}(t_0))U_\leftarrow(t,t_0) U_\rightarrow(t_0,t) B(\bs{\Gamma}(t_0))U_\leftarrow(t,t_0) 
\nonumber\\
&=& A(\bs{\Gamma}(t;t_0))\cdot B(\bs{\Gamma}(t;t_0)) ,
\end{eqnarray}
which is the end of the proof of Eq.(\ref{a_UAB=UA UB}).


\begin{thebibliography}{99}
%%%%%%%%%%%%%%%%%%%%%%%%%%%%%%%%%%%%%%%%%%%%%%%%%%%%%%%%%%%%%
% Some macros are available for the bibliography:
%  o for general use
%    \JL : general journals                 \andvol : Vol (Year) Page
%  o for individual journal 
%    \AJ   : Astrophys. J.           \NC         : Nuovo Cim.
%    \ANN  : Ann. of Phys.           \NPA, \NPB  : Nucl. Phys. [A,B]
%    \CMP  : Commun. Math. Phys.     \PLA, \PLB  : Phys. Lett. [A,B]
%    \IJMP : Int. J. Mod. Phys.      \PRA - \PRE : Phys. Rev. [A-E]     
%    \JHEP : J. High Energy Phys.    \PRL        : Phys. Rev. Lett.
%    \JMP  : J. Math. Phys.          \PRP        : Phys. Rep.
%    \JP   : J. of Phys.             \PTP        : Prog. Theor. Phys.     
%    \JPSJ : J. Phys. Soc. Jpn.      \PTPS       : Prog. Theor. Phys. Suppl.
% Usage:
%  \PRD{45,1990,345}          ==> Phys.~Rev.\ \textbf{D45} (1990), 345
%  \JL{Nature,418,2002,123}   ==> Nature \textbf{418} (2002), 123
%  \andvol{B123,1995,1020}    ==> \textbf{B123} (1995), 1020
%%%%%%%%%%%%%%%%%%%%%%%%%%%%%%%%%%%%%%%%%%%%%%%%%%%%%%%%%%%%%
  
%\bibitem{}

%\bibitem{Zubarev74}
%D.~N.~Zubarev, {\em Nonequilibrium Statistical Thermodynamics} (Consultants
%  Bureau, New York, 1974).

%\bibitem{McLennan88}
%J.~A.~McLennan, {\em Introduction to Nonequilibrium Statistical Mechanics}
%  (Prentice Hall, NJ, 1988).

%\bibitem{Sasa06}
%S.~Sasa and H.~Tasaki, \JL{J.~Stat.~Phys.,125,2006,125}

\bibitem{Evans08}
D.~J. Evans and G.~P. Morriss, {\em Statistical Mechanics of Nonequilibrium
  Liquids}, 2nd ed. (Cambridge University Press, Cambridge, 2008).

\bibitem{Morriss87}
G.~P. Morriss and D.~J. Evans, Phys. Rev. A, Application of transient correlation functions to shear flow far from equilibrium, \textbf{35} (1987) 792-797 .

\bibitem{FT93}
D.~J.~Evans, E.~G.~D.~Cohen and G.~P.~Morriss, Probability of second law violations in shearing steady states, Phys. Rev. Lett. \textbf{71} (1993) 2401-2404 .
\bibitem{GC95}
G.~Gallavotti and E.~G.~D.~Cohen, Dynamical Ensembles in Nonequilibrium Statistical Mechanics, Phys. Rev. Lett. \textbf{74} (1995) 2694-2697.

\bibitem{Kurchan}
J.~Kurchan, Fluctuation theorem for stochastic dynamics, J.~Phys.~A: Math.~Gen., \textbf{31} (1998) 3719-3729.

%\bibitem{Hatano-Sasa}
%T.~Hatano and S.~Sasa, Phys. Rev. Lett. {\bf 86}, 3643 (2001). 

\bibitem{Evans02}
D.~J. Evans and D.~J. Searles, The Fluctuation Theorem, Adv.~Phys., \textbf{51}  (2002) 1529-1585.

\bibitem{Seifert12}
U. Seifert, Stochastic thermodynamics, fluctuation theorems and molecular machines, Rep. Prog. Phys. {\bf 75} (2012) 126001 (1-58).

\bibitem{Jarzynski}
C.~Jarzynski, Nonequilibrium Equality for Free Energy Differences, Phys. Rev. Lett. \textbf{78}  (1997) 2690-2693.

\bibitem{Crooks}
G.~E.~Crooks, Path-ensemble averages in systems driven far from equilibrium, Phys. Rev. E \textbf{61} (2000) 2361 -2366.

\bibitem{mennon04} K. Feitosa and N. Mennon, Fluidized Granular Medium as an Instance of the Fluctuation Theorem, Phys. Rev. Lett. {\bf 92} (2004) 164301 (1-4).

\bibitem{Chong09b}
S.-H.~Chong, M.~Otsuki and H.~Hayakawa, Generalized Green-Kubo relation and integral fluctuation theorem for driven dissipative systems without microscopic time reversibility, Phys. Rev. E \textbf{81} (2010) 041130 (1-4).

\bibitem{kumar11} N. Kumar, S. Ramaswamy and A. K. Sood, Symmetry Properties of the Large-Deviation Function of the Velocity of a Self-Propelled Polar Particle, Phys. Rev. Lett. {\bf 106} (2011) 118001 (1-4).

\bibitem{joubaud12} S. Jaubaud, D. Lohse and D. van der Meer, Fluctuation Theorems for an Asymmetric Rotor in a Granular Gas, Phys. Rev. Lett. {\bf 108} (2012)  210604 (1-5).

\bibitem{naert12} A. Naert, Experimental study of work exchange with a granular gas: The viewpoint of the Fluctuation Theorem, EPL {\bf 97}  (2012) 20010 (1-6).

\bibitem{mounier12} A. Mounier and A. Naert, The Hatano-Sasa equality: Transitions between steady states in a granular gas,  EPL {\bf 100} (2012) 30002 (1-7).

\bibitem{Puglisi05} A. Puglisi, P. Visco, R. Barrat, E. Trizac and F. van Wijland, Fluctuations of Internal Energy Flow in a Vibrated Granular Gas, Phys. Rev. Lett. {\bf 95} (2005) 110202 (1-4).

\bibitem{Puglisi05EPL}  A. Puglisi, P. Visco, E. Trizac and F. van Wijland, Injected power and entropy flow in a heated granular gas, EPL {\bf 72} (2005) 55-61.

\bibitem{Puglisi06PRE} A. Puglisi, P. Visco, E. Trizac and F. van Wijland, Dynamics of a tracer granular particle as a nonequilibrium Markov process, Phys. Rev. E {\bf 73} (2006) 021301 (1-13).

\bibitem{Puglisi06JSM} A. Puglisi, L. Rondoni, and A. Vulpiani, Relevance of initial and final conditions for the fluctuation relation in Markov processes, J. Stat. Mech. (2006) P08001 (1-22).

\bibitem{Sarracino10} A. Sarracino, D. Villamaina, G. Gradenigo and A. Puglisi, Irreversible dynamics of a massive intruder in dense granular fluids, EPL {\bf 92} (2010) 34001 (1-5).

\bibitem{Evans94} D. J. Evans and D. J. Searles, Equilibrium microstates which generate second law violating steady states, Phys. Rev. E {\bf 50} (1994)  1645-1648.


\bibitem{Chong10}
S.-H.~Chong, M.~Otsuki and H.~Hayakawa, Representation of the Nonequilibrium Steady-State Distribution Function for Sheared Granular Systems,
Prog. Theor. Phys. Suppl. No.{\bf 184} (2010) 72-87.

\bibitem{Hayakawa10a}
H. Hayakawa, S.-H. Chong and M. Otsuki,
AIP Conf. Proc. {\bf 1227} (2010) 19-30.


\bibitem{HO12}
H. Hayakawa and M. Otsuki, Nonequilibrium identities and response theory for dissipative particles, 
Phys. Rev. E {\bf 88} (2013) 032117 (1-9).


%\bibitem{Jarzynski}
%C. Jarzynski, Phys. Rev. Lett. \textbf{78}, 2690 (1997).




\bibitem{HO2008} H. Hayakawa and M. Otsuki, Mode-Coupling Theory of Sheared Dense Granular Liquids, Prog. Theor. Phys. {\bf 119} (2008)  381-402.
\bibitem{Chong_etal_in_preparation}  K. Suzuki, S.-H. Chong, M. Otsuki and H. Hayakawa, in preparation.

\bibitem{Suzuki-Hayakawa} K. Suzuki and H. Hayakawa, Nonequilibrium mode-coupling theory for uniformly sheared underdamped systems, Phys. Rev. E . {\bf 87} (2013) 012304 (1-27).


%\bibitem{Evans90}
%D.~J. Evans and G.~P. Morriss, {\em Statistical Mechanics of Nonequilibrium
%  Liquids}, 1st ed. (Cambridge University Press, Cambridge, 1990).




\bibitem{Kranz10} W. T. Kranz, M. Sperl, and  A. Zippelius, Glass Transition for Driven Granular Fluids, Phys. Rev. Lett. {\bf 104}  (2010) 225701 (1-4).
\bibitem{Kranz12}  W. T. Kranz, M. Sperl, and A. Zippelius, Glass transition in driven granular fluids: A mode-coupling approach, Phys. Rev. E {\bf 87} (2013) 022207 (1-14).
\bibitem{SH13} K. Suzuki and H. Hayakawa, Mode-coupling theory for sheared granular liquids, AIP Conf. Proc. {\bf 1542} (2013) 670-673.

\bibitem{SH13new} K. Suzuki and H. Hayakawa, Rheology of Dense Sheared Granular Liquids: a Mode-Coupling Approach, arXiv:1310.3042 v1 (1-9).

%\bibitem{otsuki07} M. Otsuki and H. Hayakawa, J. Stat. Mech: Theor. Exp. (2009) P08003.
%\bibitem{brilliantov} N. V. Brilliantov and T. P\"{o}schel,  {\it Kinetic Theory of Granular Gases} (Oxford Univ. Press, Oxford, 2004).
%\bibitem{jenkins} J. T. Jenkins and M. W. Richman, Phys. Fluids {\bf 28} (1985) 3485.
%\bibitem{garzo} V. Garz\'{o} and J. W. Dufty, Phys. Rev. E {\bf 59} (1998) 5895.
%\bibitem{lutsko05} J. F. Lutsko, Phys. Rev. E {\bf 72} (2005) 021306.
%\bibitem{saitoh} K. Saitoh and H. Hayakawa, Phys. Rev. E {\bf 75}  (2007) 021302.
%\bibitem{hh-mo07} H. Hayakawa and M. Otsuki, Phys. Rev. E {\bf 76}  (2007) 051304.

%\bibitem{evans} D. J. Evans and G. P. Morriss, {\it Statistical Mechanics of Nonequilbrium Liquids} (Academic Press, London, 1990).



%\bibitem{resibois} P. M. V. Resibois and M. de Leener, Classical kinetic theory of fluids (John Wiley $\&$ Sons, New York, 1977).

%\bibitem{sagawa2011} T. Sagawa and H. Hayakawa, Phys. Rev. E {\bf 84}, 051110 (2011).
%
%\bibitem{saitoh2011}
%K. Saitoh and H. Hayakawa, Granular Matter \textbf{13}, 697 (2011).

%\bibitem{Sela}
%N. Sela, I. Goldhirsch and S. H. Noskowicz, Phys. Fluids {\bf 8}, 2337 (1996).

%\bibitem{otsuki} M. Otsuki and H. Hayakawa, in preparation.

%\bibitem{sagawa2011}
%T. Sagawa and H. Hayakawa, Phys. Rev. E \textbf{84}, 051110 (2011).




%\bibitem{sagawa} T. Sagawa and M. Ueda, Phys. Rev. Lett. {\bf 104}, 090602 (2010).



%\bibitem{suzuki}


\end{thebibliography}
\end{document}